\documentclass{article}

\makeatletter
\newcommand{\chapterauthor}[1]{%
 {\parindent0pt\vspace*{0pt}%
 \linespread{1.0}\normalsize\scshape#1%
 \par\nobreak\vspace*{15pt}}
 \@afterheading%
}
\makeatother

\usepackage[a4paper,margin=1in]{geometry}
\usepackage{parcolumns,lipsum}
\usepackage{verbatim}
\usepackage{subcaption}
\usepackage{url}
\usepackage{xcolor}

\title{Enrico Fermi: a forgotten article written for Italian teachers}
\author{Emanuele Goldoni, Ledo Stefanini\\ \textit{Accademia Nazionale Virgiliana}\\ \textit{Mantova, Italy}}
\begin{document}
\maketitle

\begin{abstract}
In 1929, Enrico Fermi wrote ``Problemi attuali della fisica'' (``Contemporary Problems of Physics''), a short article in Italian published in the magazine ``Annali dell'istruzione media''. The magazine was sponsored by the Italian Ministry of Schooling and Education, and it was addressed to teachers and principals of middle and high schools.
This short text written by Fermi had been forgotten for a long time. However, in recent years it has been republished in Italy; moreover, the ``Annali dell'istruzione media'' have been digitized and made freely available online.
It is unclear why Fermi wrote an article of a clearly divulgative nature for a non-technical journal, but is still interesting to to read his words and appreciate his skills as a great scientific communicator.
In this work we include the transcription of the original article in Italian, and we also propose an English translation to make the text available world-wide and accessible to a broader public.
\end{abstract}

\section*{Introduction}
Most of the scientific papers written by Enrico Fermi are contained in ``Enrico Fermi: Collected Papers'' \cite{Fermi:1962:collected1}, a collection edited by some of his colleagues and friends in 1962, just a few years after his death. Almost all of his papers which are not included in this collection are either merely summaries of articles already published elsewhere or accounts of Fermi's lectures, sometimes as transcribed by others. However, the bibliography of \cite{Fermi:1962:collected1} mentions the article ``Problemi attuali della fisica'' ({\em ``Contemporary Problems of Physics} \cite{Fermi:1929:PAD} without including it. This work is a short text in Italian written by the young Fermi and published in 1929. What makes it most interesting is its publication in ``Annali dell'istruzione media'' \cite{Monnier:1926:Annali} -- a journal sponsored by the Italian Ministry of Schooling and Education. The magazine was addressed to teachers and principals of middle and high schools; its aim was to present news and reflections on the school system, at a time when fascism had not yet been consolidated as a regime. This journal also included articles on the current problems in the relevant disciplines, including sciences.

The historical context of the Italian school system at the time makes it more interesting the presence of Fermi's article into this journal. By 1923 the Gentile's school reform had come into effect in Italy.
This reform gave primacy to humanities subjects: philosophy was to be at the center of the whole school education. Hence, the Classical lyceum (``Ginnasio'') was considered the only school capable of guaranteeing an adequate preparation to the middle and ruling classes.
Paradoxically, scientific culture was penalized in a society that was just coming out of a war and where science and technology played a key role on the economic and military fronts.
In 1927, the great mathematician Federigo Enriques wrote an article on the teaching of Mathematics and Physics in the Gentile reform \cite{Enriques:1927:insegnamento}, noting how the curricula were insufficient to the educational needs -- in the Ginnasio there was only one hour per week devoted to calculus exercises. Enriques closed the same article asking to give the mathematical and physical disciplines ``the place they deserve in humanistic education, that is to say increasing the number of chairs, and of teaching hours''.

Never subsequently republished after his death, Enrico Fermi's short article ``Problemi attuali della fisica'' had long been forgotten. However, in recent years the article was included in \cite{Barone:2009:scritti}; moreover, the National Central Library in Rome began a digitization effort of its collections and the issues of ``Annali dell'istruzione media'' are now freely available online \cite{Roma:2023:bnc}.

We do not know why Fermi wrote an article of a clearly divulgative nature for a non-technical journal for teachers and principals, while he was engaged at that very time, jointly with Corbino, in transforming the Via Panisperna Institute into a world-class state-of-the-art center. Unfortunately, there is no mention of this article in Fermi's biographies \cite{FermiL:1954:atoms, Schwartz:2017:lastman}.

For sure, in those years Fermi was well aware of what was happening within the Italian school. In fact, in 1929 he published for Zanichelli a physics book for high schools \cite{Fermi:1929:fisica} in order to increase his income and supplement his modest salary as a professor of theoretical physics at the University of Rome. It is possible that Fermi simply wanted to stimulate teachers, largely unaware of the emerging quantum theory, and to raise the level of physics teaching in schools reviving interest in the subject.

Whatever Fermi's real reasons for writing this article were, it is still interesting to be able to read again his words and appreciate his skills as a great popularizer. Indeed, the text clearly shows his ability to render even very difficult concepts without making them seem so, lightening the theoretical treatment with frequent references to experimental activities.

In this work we include the transcription of the original article in Italian, and we also propose an English translation. We are aware that translating a work written by Fermi about a century ago could seem inappropriate, especially considering his well-known concern for the precise expression. Our decision was only motivated by the desire to make the text available world-wide and accessible to a broader public. Thus, we have tried to keep the style and the vocabulary as faithful to the original text as possible, although some words and sentences has been adapted to sound more familiar to a reader of the 21st century.

\section*{Problemi attuali della fisica}
\chapterauthor{Enrico Fermi}
La fisica degli ultimi decenni, sia nel campo sperimentale che in quello teorico, è stata caratterizzata da un progressivo aumento dell'interesse per i fenomeni atomici.

Tutti i fenomeni accessibili all'osservazione diretta, e gran parte di quelli osservabili con mezzi di precisione non grandissima, non sono altro che il risultato di un numero enorme di processi elementari. Così la pressione, esercitata da un gas sopra una parete, non è altro che il risultato di innumerevoli urti delle molecole del gas contro la parete; similmente quando la luce viene assorbita da una sostanza, si ha un grande numero di atti di assorbimento da parte dei singoli atomi. È evidente l'importanza dello studio di questi fenomeni elementari, che formano la chiave per comprendere i fenomeni più complessi. Naturalmente però il loro studio sperimentale è molto difficile, poiché l'elemento su cui si opera è l'atomo, oppure la molecola, entrambi di dimensioni estremamente piccole.

Cercheremo qui di passare in rapida rassegna alcune delle principali esperienze in cui si arriva all'osservazione diretta di uno di questi fenomeni elementari. 

\centerline{* * *}

Una delle prime osservazioni è stata quella dei moti Browniani delle particelle colloidali, visibili direttamente per mezzo dell'ultra-microscopio; il movimento delle particelle è dovuto, come è ben noto, agli urti che esse ricevono da parte delle molecole del liquido in cui sono immerse, le quali sono in movimento per effetto dell'agitazione termica.

Il più potente metodo per l'investigazione dei fenomeni elementari è quello della nebbia di Wilson, che permette di rendere visibili gli ioni e gli elettroni che si trovano in un ambiente.

Esso è basato sopra la proprietà dei corpuscoli elettrizzati, di servire da nuclei di condensazione del vapore acqueo. L'apparecchio di Wilson è costituito da un recipiente che contiene dell'aria umida; spostando un pistone si può produrre una rapida espansione dell'aria, la quale raffredda fino al punto che il vapore d'acqua contenuto in essa diventa soprasaturo. Se non vi fosse traccia di nuclei di condensazione nell'aria, non si avrebbe tuttavia condensazione del vapore, almeno quando si mantenga l'espansione dell'aria e il conseguente raffreddamento entro opportuni limiti. Ma se nell'aria umida si trovano degli ioni, il vapore acqueo si condensa intorno ad essi per modo che attorno ad ogni ione viene a formarsi una gocciolina d'acqua che lo rende visibile.

Una delle più note applicazioni di questo metodo di Wilson si ha nella osservazione delle traiettorie delle particelle $\alpha$. Quanto una particella $\alpha$ viene proiettata con grandissima velocità attraverso all'aria da una sostanza radioattiva, essa spezza le molecole che incontra sul suo cammino, lasciando sul suo percorso un grande numero di ioni, residuo delle molecole spezzate. Poniamo entro una camera di Wilson un piccolo granulo di sostanza radioattiva, che proietterà tutto intorno a sé delle particelle $\alpha$. Supponiamo poi di produrre l'espansione dell'aria umida della camera, un momento dopo che essa è stata traversata da una particella. Il vapore acqueo si condenserà allora sugli ioni restati lungo la traiettoria della particella $\alpha$, per modo che lungo di essa si verrà a formare una serie di goccioline di nebbia bene visibili, specialmente se illuminate fortemente di traverso. Con lo stesso metodo si possono rendere visibili anche le traiettorie degli elettroni, benché la loro osservazione sia di solito un po' meno precisa, poiché gli elettroni producono sul loro passaggio un numero di ioni notevolmente minore delle particelle $\alpha$.

Il metodo della nebbia per la osservazione delle traiettorie delle particelle $\alpha$ è stato elaborato specialmente nella scuola di Rutherford a Cambridge e ha portato a risultati di eccezionale interesse. Uno dei più importanti consiste nell'osservazione della così detta disintegrazione artificiale dei nuclei di certi atomi. Quando una particella $\alpha$ urta contro il nucleo di un atomo, può accadere che essa lo spezzi strappandone un protone, il quale non è altro che un nucleo di idrogeno; la particella urtante può rimbalzare via, oppure restare aggregata al nucleo urtato. Così per es. supponiamo di bombardare dell'azoto con delle particelle $\alpha$ molto veloci. L'azoto ha numero atomico 7, cioè la carica elettrica del suo nucleo è sette volte la carica elettrica elementare del protone. Se una particella $\alpha$ urta un nucleo di azoto e ne distacca un protone, la carica del nucleo residuo resta 6; supponendo che la particella $\alpha$ urtante (di carica 2) resti aggregata al nucleo, la carica elettrica del complesso che ne risulta sarà 8. Ora l'elemento chimico il cui nucleo ha carica 8 è l'ossigeno; come risultato del processo di urto si verrà quindi ad avere produzione di ossigeno. Disgraziatamente la probabilità che una particella $\alpha$, urtando un nucleo di azoto, lo spezzi nel modo che è stato descritto, è estremamente piccola, per modo che, per quanto intensamente si bombardi l'azoto con le particelle $\alpha$, la quantità di ossigeno prodotta resta sempre così piccola, da non essere nemmeno lontanamente osservabile con metodi chimici.

Per scoprire la formazione dell'ossigeno si può invece ricorrere appunto al metodo di Wilson. Occorre perciò esaminare qualche decina di migliaia di fotografie di traiettorie di particelle $\alpha$ nell'azoto; si riesce così a scoprire qualche traiettoria che, ad un certo punto, si biforca. Uno dei due rami è la traccia del passaggio del protone strappato, per effetto dell'urto, dal nucleo di azoto; l'altro ramo, assai più corto, è invece prodotto dal passaggio del nucleo residuo, al quale si è aggregata la particella $\alpha$ urtante. Se dopo l'urto la particella $\alpha$ fosse invece rimasta libera, si sarebbe osservata una traccia che, al momento dell'urto, si doveva dividere in tre rami.

Anche fuori del campo della radioattività sono state fatte delle applicazioni del metodo di Wilson di grandissimo interesse. Una delle più significative è l'esperienza di Compton e Simon che illustra in modo notevolissimo il meccanismo dell'effetto Compton. Secondo la teoria dei quanti di luce, il processo della diffusione dei raggi X per parte degli atomi (effetto Compton) è il seguente: Consideriamo un quanto di luce di frequenza $\nu$ che urti contro un elettrone. Esso verrà diffuso, cambiando così di direzione; siccome però un quanto di luce, oltre alla sua energia $h\nu$, ha anche una quantità di moto, l'elettrone che ha prodotto il cambiamento di direzione del quanto, riceverà un impulso, eguale alla differenza vettoriale delle quantità dei moto del quanto di luce, prima e dopo l'urto. L'elettrone acquista così, in seguito all'urto, una certa velocità, e l'energia cinetica corrispondente viene fornita a spese del quanto di luce, la cui energia diminuisce per ciò nell'urto. Siccome poi la frequenza di un quanto è proporzionale alla sua energia, la frequenza dei raggi diffusi risulterà un po' minore della frequenza di quelli incidenti. L'esperienza verifica in realtà questa diminuzione della frequenza, e la diminuzione osservata corrisponde anche molto bene al calcolo teorico. Aveva tuttavia interesse la verifica diretta del meccanismo che abbiamo descritto; si trattava cioè di far vedere che effettivamente tutte le volte che un quanto di luce viene diffuso in una data direzione, viene in corrispondenza proiettato un elettrone in una direzione che può facilmente prevedersi calcolando, per mezzo dei principi di conservazione dell'energia, e della quantità di moto, il processo dell'urto tra il quanto di luce e l'elettrone. Questa verifica è stata fatta da Compton e Simon col metodo seguente.

In una camera di Wilson essi mandarono un sottile fascetto di raggi X e fecero poi qualche migliaio di fotografie delle tracce di nebbia che si formavano nella camera, producendo delle espansioni. Ogni volta venivano prese contemporaneamente due fotografie in direzioni perpendicolari, in modo da poter determinare stereoscopicamente la direzione delle tracce. In alcune delle fotografie si osservano delle tracce dovute a un elettrone che è rimbalzato per effetto della diffusione di un quanto. La direzione della porzione iniziale della traccia ci dà la direzione in cui è stato proiettato l'elettrone per effetto del rimbalzo; da questa si può calcolare, con la teoria dell'urto, la direzione in cui è stato diffuso il quanto urtante. Ora accade abbastanza spesso che il quanto diffuso, nell'attraversare il gas della camera, venga assorbito da un atomo, e gli strappi un elettrone per effetto fotoelettrico. In questo caso noi vedremo sulla stessa fotografia, insieme alla traccia dell'elettrone di rimbalzo, anche un'altra traccia che incomincia da un punto della traiettoria del quanto diffuso. Compton e Simon riuscirono effettivamente a osservare in un buon numero di fotografie delle coppie di tracce disposte nel modo che abbiamo descritto.

Il risultato di questa esperienza si accorda perfettamente con le previsioni della teoria dei quanti di luce. Essa fu una delle più tipiche esperienze che servirono a precisare il contrasto tra questa teoria e quella ondulatoria; contrasto che ha oggi trovato la sua soluzione nella nuova meccanica quantistica. 

\centerline{* * *}

Ho cercato, con questi pochi esempi, di illustrare l'importanza che hanno avuto le esperienze sui fenomeni elementari nello sviluppo della fisica moderna. Molto però ci si può ancora attendere da esperienze di questo tipo per chiarire quei problemi, la cui risoluzione costituirà il compito della fisica di domani. Tra questi forse il più importante è quello della causalità. Nella fisica dei fenomeni macroscopici, tutti gli stati futuri di un sistema sottratto a qualsiasi perturbazione esterna sono determinati univocamente dalla conoscenza dello stato iniziale del sistema stesso (determinismo). Si hanno invece varie ragioni che sembrano indicare che un simile principio deterministico non sia valido nel mondo microscopico degli atomi. L'apparente determinismo macroscopico sarebbe unicamente il risultato del fatto che, nelle osservazioni macroscopiche, si osservano soltanto delle medie di numerosissimi fenomeni che avvengono negli atomi costituenti i diversi corpi. Non sfuggirà a nessuno l'enorme importanza di questa questione e il profondo mutamento che avverrebbe nelle nostre vedute sopra i fenomeni naturali, il giorno che ci si dovesse effettivamente convincere che il determinismo fisico è soltanto una proprietà di media, non più valida quando si opera sopra un singolo atomo. È appunto da uno studio profondo dei fenomeni elementari che si potrà avere la risposta a questo appassionante problema; si potrà vedere cioè se l'apparente indeterminismo dei fenomeni elementari derivi dal fatto che si è omessa l'osservazione di qualche causa, oppure se esso rappresenti una delle leggi fondamentali del mondo atomico.

\section*{Contemporary Problems of Physics}
\chapterauthor{Enrico Fermi (Author), Emanuele Goldoni and Ledo Stefanini (Translators)}
The physics of recent decades, both in the experimental and the theoretical fields, has been characterized by a increasing interest in atomic phenomena.

All phenomena observable directly, and most of those observable through not-so-precise means, are nothing but the result of an enormous amount of elementary processes. Thus, the pressure exerted by a gas over a wall, is just the result of countless collisions of gas molecules against the wall; similarly, when light is absorbed by a substance, there is a great number of events of absorption by individual atoms. The importance of studying these elementary phenomena - which are the key to understand more complex phenomena - is obvious. However, their experimental investigation is very difficult, since one must work with either the atom or the molecule, both of which are extremely small.

Here, we will try to quickly review some of the main experiences in which we arrive at a direct observation of one of these elementary phenomena. 

\centerline{* * *}

One of the earliest observations involved the Brownian motions of colloid particles, which are visible directly by means of an ultramicroscope. It is well known that the movement of the particles is due to the shocks they receive from the molecules of the liquid in which they are immersed, which are in motion due to thermal agitation.

The most powerful method for investigating elementary phenomena is the Wilson cloud chamber, which makes visible the ions and electrons found in an environment.

The Wilson chamber is based on the property of electrified corpuscles, which act as condensation nuclei of water vapor. It consists of a device containing moist air -- moving a piston, we can produce a rapid expansion of the air, which cools to the point that the contained water vapor becomes supersaturated. However, if there were no trace of condensation nuclei in the air, there would be no condensation of vapor, assuming the expansion of the air and the consequent cooling are kept within suitable limits. On the contrary, if there are ions in the moist air, water vapor condenses around them so that a water droplet is formed around each ion, making it visible.

One of the best known applications of Wilson's device is the observation of the trajectories of $\alpha$ particles. When an $\alpha$ particle is thrown at a very high velocity through the air by a radioactive substance, it breaks up the molecules it encounters on its path; thus, it leaves on its path a large number of ions, which are the residue of the broken molecules. Let's place a small granule of radioactive substance within a Wilson chamber: it will throw $\alpha$ particles all around. Then, let's suppose that we generate the expansion of the moist air of the chamber just after it has been traversed by a particle. The water vapor will then condense on the ions left along the trajectory of the $\alpha$ particle: hence, a series of mist droplets will form and it will be well visible, especially when tangentially strongly illuminated. The trajectories of the electrons could be made visible by the same method, although their observation is usually somewhat less precise, since electrons produce considerably fewer ions than $\alpha$ particles on their passage.

The cloud method for observing the trajectories of $\alpha$ particles was mostly developed by the Rutherford's school in Cambridge and has led to results of exceptional interest. One of the most important achievements consists in the observation of the so-called artificial disintegration of the nuclei of certain atoms. When an $\alpha$ particle collides against the nucleus of an atom, it may happen that it breaks it up by tearing off a proton, which is nothing but an hydrogen nucleus. Then, the colliding particle might bounce away or be captured by the hit nucleus. Let's suppose, for example, we bombard nitrogen with very fast $\alpha$ particles. Nitrogen has atomic number 7, that is, the electric charge of its nucleus is seven times the elemental electric charge of the proton. If an $\alpha$ particle collides with a nitrogen nucleus and detaches a proton from it, the charge of the remaining nucleus remains 6. Assuming instead that the bumping $\alpha$ particle (of charge +2) remains aggregated to the nucleus, the electrical charge of the resulting complex will be 8. Since the chemical element whose nucleus has charge 8 is oxygen, the collision process will therefore produce oxygen. Unfortunately, the probability that an $\alpha$ particle, colliding with a nitrogen nucleus, would break it up in the manner that has been just described, is extremely small -- no matter how intensely one bombards nitrogen with $\alpha$ particles, the amount of oxygen produced will always remains so small that it is not even remotely observable by chemical methods.

To discover the formation of oxygen, one can instead turn to Wilson's method. It is therefore necessary to examine a few tens of thousands photographs of trajectories of $\alpha$ particles inside nitrogen; one is thus able to discover a few trajectories that, at some point, bifurcate. One of the two branches is the trace of the passage of the proton, ejected by the impact from the nitrogen nucleus; the other branch, which is much shorter, is produced instead by the passage of the recoil nucleus, to which the impacting $\alpha$ particle has aggregated to. If after the collision the $\alpha$ particle had instead remained free, one would have observed a track which, when collision took place, should have split into three branches.

There have been very interesting applications of Wilson's method beyond radioactivity, too. One of the most significant applications is the experience of Compton and Simon, which remarkably illustrates the mechanism of the Compton effect. According to the quantum theory of light, the process of X-ray scattering by atoms (Compton effect) is as follows: consider a quantum of light of frequency $\nu$ colliding with an electron. The quantum will be scattered, thus changing direction. However, a light quantum, in addition to its energy $h\nu$, also has momentum: then, the electron that produced the change in direction of the quantum will receive an impulse, equal to the vector difference of the momentum of the light quantum before and after the collision. The electron will thus acquire a certain velocity, as a result of the collision; the corresponding kinetic energy is supplied at the expense of the quantum of light, whose energy therefore decreases in the collision. Moreover, since the frequency of a quantum is proportional to its energy, the frequency of the scattered rays will be somewhat less than the frequency of the incident ones. Experience actually confirms this decrease in frequency, and the observed decrease also corresponds very well to the theoretical calculation. However, there was interest in verifying directly the mechanism we have described; that is, to show that every time a quantum of light is scattered in a given direction, a corresponding electron is ejected in a predictable direction. The direction of the electron could be easily calculated, by means of the principles of conservation of energy and momentum, considering the effect of the collision between the quantum of light and the electron. This verification was done by Compton and Simon by the following method.

In a Wilson chamber, they sent a thin beam of X-rays; then, they produced expansions and took a few thousand photographs of the cloud tracks forming inside the chamber. Each time, two photos were taken simultaneously in perpendicular directions, so that the direction of the tracks could be determined stereoscopically. In some of the photos, traces were observed due to an electron that had bounced back as a result of quantum scattering. The direction of the initial portion of the trace gives us the direction in which the electron was projected due to the bounce; from this, we can calculate the direction in which the colliding quantum was diffused via the collision theory. It happens quite often that the scattered quantum, in passing through the gas within the chamber, is absorbed by an atom and removes an electron from the same atom by the photoelectric effect. In this case, we will see on the same photograph the trace of the bouncing electron plus another trace beginning at a point in the trajectory of the scattered quantum. Compton and Simon actually succeeded in observing pairs of tracks, arranged as we have just described, in a good number of photographs.

The result of this experience agrees perfectly with the predictions of the quantum theory of light. It was one of the most typical experiences that served to clarify the contrast between this theory and the wave theory; a contrast that has now found its solution in the new quantum mechanics. 

\centerline{* * *}

With these few examples, I tried to illustrate the importance that experiments on elementary phenomena have had in the development of modern physics. However, much more can still be expected from such experiences in order to clarify those problems whose resolution will be the task of tomorrow's physics. Among these problems, perhaps the most important is that of causality. In the physics of macroscopic phenomena, all future states of a system removed from any external perturbation are uniquely determined by knowledge of the system's initial state (determinism). There are, however, various reasons that seem to indicate that such a deterministic principle does not hold in the microscopic world of atoms. The apparent macroscopic determinism would be solely the result of the fact that, in macroscopic observations, we observe only averages of many phenomena occurring in the constituent atoms of bodies. The enormous importance of this question is evident, and a profound change would take place in our view over natural phenomena the day that we should actually be convinced that physical determinism is only a property of averages, no longer valid when operating on a single atom. It is precisely from a thorough study of elementary phenomena that the answer to this exciting problem may come; that is, we will be able to verify if the apparent indeterminism of elementary phenomena is due to the missing observation of some causes or it represents one of the fundamental laws of the atomic world.

\section*{Acknowledgements}
We would like to thank Davide Cavalca and Alessandro Vezzani for giving the paper a critical reading and for providing several helpful comments.

\end{document}